\documentclass{appolb}
\usepackage{graphicx}

\providecommand\eprint[1]{\href{http://arXiv.org/abs/#1}{arXiv:#1}}

\usepackage[numbers,sort&compress]{natbib}

\usepackage{hyperref}
\hypersetup{
  breaklinks=false,
}

\usepackage{amsmath}

\usepackage{etoolbox}

\newtoggle{prerefAA} \newtoggle{prerefBB} \newtoggle{prerefCC}

\togglefalse{prerefAA}
\togglefalse{prerefBB}
\togglefalse{prerefCC}

\iftoggle{prerefAA}{
  \newcommand\prerefereeAAchanges[1]{{\bf \large \color{myred} #1}}     \usepackage{color}  \definecolor{myred}{rgb}{0.7,0.0,0.2}
}{
  \newcommand\prerefereeAAchanges[1]{#1}    
}

\iftoggle{prerefBB}{
  \newcommand\prerefereeBBchanges[1]{{\bf \large \color{myred} #1}}     \usepackage{color}  \definecolor{myred}{rgb}{0.7,0.0,0.2}
}{
  \newcommand\prerefereeBBchanges[1]{#1}    
}

\iftoggle{prerefCC}{
  \newcommand\prerefereeCCchanges[1]{{\bf \large \color{myred} #1}}     \usepackage{color}  \definecolor{myred}{rgb}{0.7,0.0,0.2}
}{
  \newcommand\prerefereeCCchanges[1]{#1}    
}

\newcommand\Ommzeroeff{\Omega_{\mathrm{m}0}^{\mathrm{eff}}}
\newcommand\OmQzeroeff{\Omega_{\CQ0}^{\mathrm{eff}}}

\newcommand\OmRzeroeff{\Omega_{{\cal R}0}^{\mathrm{eff}}}

\newcommand\aeff{a_{\mathrm{eff}}}

\newcommand\abg{a_{\mathrm{bg}}}
\newcommand\abgzero{a_{\mathrm{bg0}}}

\newcommand\Hbg{H_{\mathrm{bg}}}
\newcommand\dotabg{\dot{a}_{\mathrm{bg}}}

\newcommand\Hzeroeff{H_0^{\mathrm{eff}}}
\newcommand\Hzerobg{H_0^{\mathrm{bg}}}
\newcommand\Honebg{\mbox{{$H_{1}^{\mathrm{bg}}$}}}

\newcommand\tzeroefffull{{t_{\aeff=1}}}
\newcommand\tzeroeff{{t_0}}

\newcommand\Hpeculiarzero{H_{\mathrm{pec},0}^{\mathrm{void}}}

\newcommand{\initial}{^{\prerefereeBBchanges{\mathrm{init}}}}

\providecommand\kmsMpc{\mbox{\,$\mathrm{km/s/Mpc}$}}

\providecommand\lsim{\mathop{\hbox{${\lower3.8pt\hbox{$<$}}\atop{\raise0.2pt\hbox{$\sim$}}
$}}}
\providecommand\gsim{\mathop{\hbox{${\lower3.8pt\hbox{$>$}}\atop{\raise0.2pt\hbox{$
          \sim$}}$}}}

\providecommand\colorrgb[1]{\color[rgb]{#1}} 
\providecommand\colorgray[1]{\color[gray]{#1}} 
\providecommand\rotatebox[2]{#2}

\providecommand\jcap{Journ.~Cosm.~Astr.~Phys.}
\providecommand\prd{Phys.~Rev.~D}

\providecommand\mnras{Mon.~Not.~Roy.~Astr.~Soc.\/}
\providecommand\aap{Astron.~Astroph.\/}

\providecommand\cqg{Class.~Quant.~Gra.\/}   %
\providecommand\grg{Gen.~Rel.~Grav.\/}
\providecommand\annrevnucpartphys{Ann.~Rev.~Nucl.~Part.~Sci.}

\providecommand\newjphys{New Journ.~Phys.\/}

\newcommand{\CQ}{{\cal Q}}

\begin{document}
\title{Order-unity argument for structure-generated ``extra'' expansion%
  \thanks{Presented at the 3rd Conference of the Polish Society on Relativity.}}
\author{Boudewijn F. Roukema$^1$,
  Jan J. Ostrowski$^2$,
  Thomas Buchert$^2$,
  Pierre Mourier$^2$
\address{$^1$Toru\'n Centre for Astronomy,
  Faculty of Physics, Astronomy and Informatics,
  Grudziadzka 5,
  Nicolaus Copernicus University, ul. Gagarina 11, 87-100 Toru\'n, Poland}
\address{$^2$Univ Lyon, Ens de Lyon, Univ Lyon1, CNRS, Centre de Recherche Astrophysique de
  Lyon UMR5574, F--69007, Lyon, France}
}
\headtitle{Order-unity argument for structure-generated ``extra'' expansion}
\headauthor{Roukema, Ostrowski, Buchert \& Mourier}

\maketitle
\begin{abstract}
  Self-consistent treatment of cosmological structure formation and
  expansion within the context of classical general relativity may
  lead to
  \protect\prerefereeAAchanges{``extra'' expansion above that expected in a}
  structureless universe.
  \protect\prerefereeCCchanges{We argue that in comparison to an
  early-epoch, extrapolated Einstein--de~Sitter model, about 10--15\%}
  ``extra'' expansion is sufficient at the present to
  render 
  \protect\prerefereeCCchanges{superfluous the ``dark energy''
    68\% contribution to the energy density budget},
  and that this is observationally realistic.
\end{abstract}
%

\section{Introduction}
\prerefereeBBchanges{In contrast}
to Friedmann--Lema\^{\i}tre--Robertson--Walker (FLRW) cosmological models,
inhomogeneous curvature and inhomogeneous expansion
\prerefereeBBchanges{in an initially FLRW model}
can be taken into account
relativistically by using the
\prerefereeAAchanges{spatially averaged}
Raychaudhuri equation \prerefereeAAchanges{and} Hamiltonian
constraint
\cite{Buch00scalav,Rasanen04,Buchert08status,WiegBuch10,BuchRas12}, where
we write the latter
\prerefereeAAchanges{\cite[][Eq.~(41)]{Buchert08status}}
at the current epoch
\begin{equation}
  \OmRzeroeff = 1 - \Ommzeroeff - \OmQzeroeff \;,
  \label{e-curv-eqn}
\end{equation}
where $\OmRzeroeff,$ $\Ommzeroeff,$ and $\OmQzeroeff$ are the
effective (averaged) present-day scalar (3-Ricci)
\prerefereeBBchanges{curvature,
matter density, and kinematical backreaction,}
respectively,
\prerefereeBBchanges{appropriately normalised by the expansion rate squared,}
and we assume zero dark energy.
\prerefereeAAchanges{The} recent emergence of average
negative scalar curvature
($\OmRzeroeff > 0$)
in tight coupling with kinematical backreaction
may lead to \prerefereeAAchanges{an} effective
\prerefereeBBchanges{scale factor}
$\aeff(t,P_k\initial)$,
where $P_k\initial$ is the initial power spectrum of density fluctuations,
that avoids the need to introduce non-zero dark
energy when matching FLRW models to observations
(\cite{Buch00scalav, Buchert05CQGLetter, BuchCarf08,
  Buchert08status}; cf \cite{Wiltshire07clocks}).

\section{Early-epoch, extrapolated Einstein--de~Sitter ``background''}
We adopt an
early-epoch Einstein--de~Sitter (EdS) ``background'' model that we
extrapolate to the present, with scale factor $\abg$ and expansion rate
$\Hbg$ \prerefereeAAchanges{given by}
\begin{align}
  \abg := (3 \Honebg t /2)^{2/3} \,, \quad
  \Hbg := \dotabg/\abg = 2/(3t)\;,
  \label{e-defn-Hbg1}
\end{align}
where the early-epoch--normalised EdS Hubble constant
$\Honebg = 37.7\pm 0.4${\kmsMpc} is estimated by using
the Planck 2015 calibration
\prerefereeAAchanges{\cite[][Table 4, sixth data column]{Planck2015cosmparam}}
as a phenomenological proxy for
many observational datasets
\prerefereeAAchanges{\cite[][Eq.~(11)]{RMBO16Hbg1}}.
For the effective scale factor
to be observationally realistic, it would need to satisfy
$\aeff \approx \abg$ at early times $t \ll \tzeroeff$
\prerefereeAAchanges{and reach unity} at the present
$\tzeroeff \equiv \tzeroefffull$.
We assume bi-domain scalar averaging
\prerefereeAAchanges{\cite{WiegBuch10,Wiltshire07clocks}}
and virialisation of collapsed (overdense) regions
\prerefereeAAchanges{(stable clustering)}.
\prerefereeAAchanges{We} define a
\prerefereeAAchanges{present-day} background Hubble constant
\begin{align}
  \Hzerobg := \Hbg(\aeff=1)\,
  \label{e-defn-Hbg0}
\end{align}
and our stable clustering assumption leads to
\prerefereeAAchanges{\cite[][Eq.~(2.27)]{ROB13}}
\begin{align}
  \Hzeroeff \approx \Hzerobg + \Hpeculiarzero \;,
  \label{e-expansion-eqn-now}
\end{align}
where
$\Hzeroeff$ is the locally observed Hubble constant and
$\Hpeculiarzero$ is the
\prerefereeAAchanges{present-day}
peculiar expansion rate of
underdense regions, i.e., typically that of voids,
\prerefereeAAchanges{above that of}
the extrapolated background model
(\prerefereeAAchanges{not} a locally fit mean model).

The three Hubble constants can be related to
one another thanks to matter conservation and the
above equations
\prerefereeAAchanges{\cite[][Eqs~(7), (10)]{RMBO16Hbg1}}:
\begin{align}
  \Hzerobg =
  \Hzeroeff \sqrt{\Ommzeroeff {/\abgzero^3}} \;,\quad\quad
  \Honebg = \Hzeroeff \sqrt{\Ommzeroeff}\;,
\end{align}
and to the present age of the Universe via the EdS relation
following from Eq.~(\ref{e-defn-Hbg1}), i.e.
$\Hzerobg = 2/(3\tzeroeff).$

\begin{figure}[htb]
  \centering
  \input{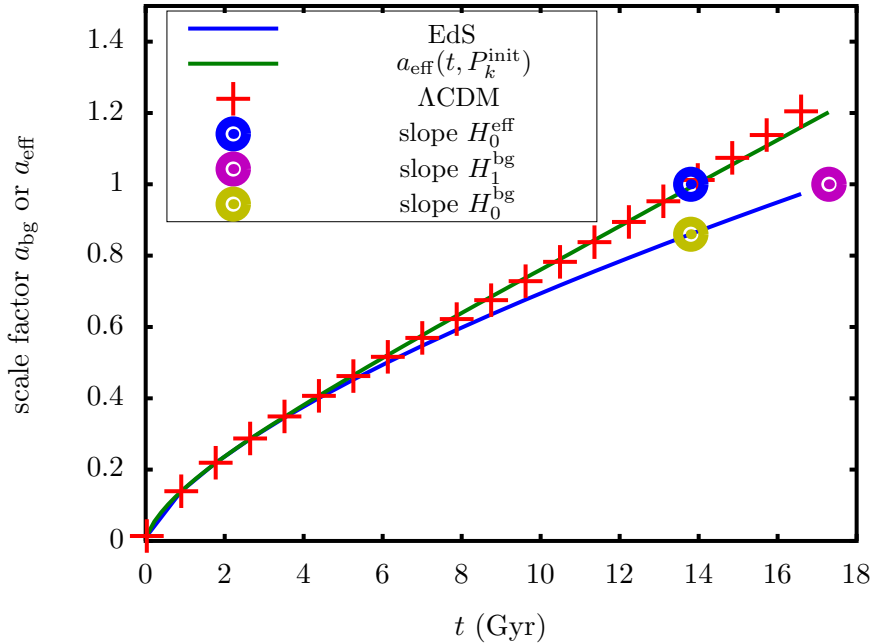}
  \caption{Observationally required Hubble constants and
    \prerefereeAAchanges{required}
    relation of \prerefereeAAchanges{the}
    effective scale factor $\aeff(t,P_k\initial)$ (upper curve) to
    \prerefereeAAchanges{the}
    background EdS scale factor $\abg(t)$ (lower curve).
    The observational proxy ($\Lambda$CDM model)
    is shown as $+$ symbols. The left and right thick circular symbols
    at unity scale factor
    correspond to normalised slopes which are the
    locally estimated $\Hzeroeff$ at $(t=13.8\,\mathrm{Gyr}, \aeff=1)$
    and the background $\Honebg$ at $(t=17.3\,\mathrm{Gyr}, \abg=1)$, while
    $\Hzerobg$ is the slope at $(t=13.8\,\mathrm{Gyr}, \abgzero = 0.86)$
    (blue, purple, yellow, respectively, online).
  }
  \label{f-aeff-needed}
\end{figure}

\section{Observational challenge}

\prerefereeAAchanges{The above definitions and equations show that
  there is very little observational parameter freedom in this class
  of cosmological backreaction models.}
The Planck~2015
observational proxy $\tzeroeff = 13.80\pm0.02$~Gyr gives
$ \Hzerobg = 47.24 \pm 0.07${\kmsMpc}, yielding a
present-day background scale factor of
\begin{align}
  \abgzero =
  \left({\Honebg/{\Hzerobg}}\right)^{2/3} =
  0.860 \pm 0.007,
\end{align}
while microlensed Galactic bulge stars give a less FLRW-model--dependent
estimate of $\abgzero = 0.90 \pm 0.01$
\cite{Bensby13Bulgetzero,RMBO16Hbg1}.

\section{Conclusion}
As \prerefereeCCchanges{shown} in Fig.~\ref{f-aeff-needed}, \prerefereeCCchanges{only} 10--15\%
``extra'' expansion
\prerefereeCCchanges{\cite[cf.][]{rasanenFOCUS}}
is needed \prerefereeBBchanges{above that of} the EdS background in order
for a dark-energy-free cosmological backreaction model to
fully replace
\prerefereeCCchanges{the ``dark energy'' 68\%} contribution to the energy density budget, i.e.
to provide an order-unity level, non-exotic alternative.
The rough observational estimate of the void peculiar expansion rate
\cite{ROB13},
and the detected Sloan Digital Sky Survey
environmental dependence of the baryon acoustic oscillation
peak scale
\cite{RBOF15,RBFO15}
provide tentative observational support for the required $\Hzeroeff - \Hzerobg$,
and $\abgzero$, respectively.

\medskip

Some of this work was supported by grant 2014/13/B/ST9/00845 of the
National Science Centre, Poland, and calculations
by Pozna\'n Supercomputing and
Networking Center grant 197.
The work of TB and JJO was conducted
under grant ANR-10-LABX-66 within the ``Lyon Institute of Origins''.

\let\tmpthebibliography\thebibliography
\renewcommand\thebibliography[1]{
  \tmpthebibliography{#1}
  \setlength{\parskip}{0ex plus 0.1ex}
  \setlength{\itemsep}{0.5ex plus 0.1ex}
}


\vspace{-3ex}
\begin{thebibliography}{15}
\expandafter\ifx\csname natexlab\endcsname\relax\def\natexlab#1{#1}\fi

\bibitem[{{Buchert}(2000)}]{Buch00scalav}
T.~{Buchert}. 2000, \grg, 32, 105, \eprint{gr-qc/9906015}

\bibitem[{{R{\"a}s{\"a}nen}(2004)}]{Rasanen04}
S.~{R{\"a}s{\"a}nen}. 2004, \jcap, 2, 003, \eprint{astro-ph/0311257}

\bibitem[{{Buchert}(2008)}]{Buchert08status}
T.~{Buchert}. 2008, \grg, 40, 467, \eprint{0707.2153}

\bibitem[{{Wiegand} \& {Buchert}(2010)}]{WiegBuch10}
A.~{Wiegand} \& T.~{Buchert}. 2010, \prd, 82, 023523, \eprint{1002.3912}

\bibitem[{{Buchert} \& {R{\"a}s{\"a}nen}(2012)}]{BuchRas12}
T.~{Buchert} \& S.~{R{\"a}s{\"a}nen}. 2012, \annrevnucpartphys, 62, 57,
  \eprint{1112.5335}

\bibitem[{{Buchert}(2005)}]{Buchert05CQGLetter}
T.~{Buchert}. 2005, \cqg, 22, L113, \eprint{gr-qc/0507028}

\bibitem[{{Buchert} \& {Carfora}(2008)}]{BuchCarf08}
T.~{Buchert} \& M.~{Carfora}. 2008, \cqg, 25, 195001, \eprint{0803.1401}

\bibitem[{{Wiltshire}(2007)}]{Wiltshire07clocks}
D.L. {Wiltshire}. 2007, \newjphys, 9, 377, \eprint{gr-qc/0702082}

\bibitem[{{Ade}  {\em et~al.}(2016)}]{Planck2015cosmparam}
P.A.R. {Ade}  {\em et~al.} 2016, \aap, 594, A13, \eprint{1502.01589}

\bibitem[{{Roukema} {et~al.}(2017){Roukema}, {Mourier}, {Buchert}, \&
  {Ostrowski}}]{RMBO16Hbg1}
B.F. {Roukema}, P.~{Mourier}, T.~{Buchert}, \& J.J. {Ostrowski}. 2017, \aap,
  598, A111, \eprint{1608.06004}

\bibitem[{{Roukema} {et~al.}(2013){Roukema}, {Ostrowski}, \& {Buchert}}]{ROB13}
B.F. {Roukema}, J.J. {Ostrowski}, \& T.~{Buchert}. 2013, \jcap, 10, 043,
  \eprint{1303.4444}

\bibitem[{{Bensby}  {\em et~al.}(2013)}]{Bensby13Bulgetzero}
T.~{Bensby}  {\em et~al.} 2013, \aap, 549, A147, \eprint{1211.6848}

\bibitem[{{R{\"a}s{\"a}nen}(2011)}]{rasanenFOCUS}
S.~{R{\"a}s{\"a}nen}. 2011, \cqg, 28, 164008, \eprint{1102.0408}

\bibitem[{{Roukema} {et~al.}(2015){Roukema}, {Buchert}, {Ostrowski}, \&
  {France}}]{RBOF15}
B.F. {Roukema}, T.~{Buchert}, J.J. {Ostrowski}, \& M.J. {France}. 2015,
  \mnras, 448, 1660, \eprint{1410.1687}

\bibitem[{{Roukema} {et~al.}(2016){Roukema}, {Buchert}, {Fujii}, \&
  {Ostrowski}}]{RBFO15}
B.F. {Roukema}, T.~{Buchert}, H.~{Fujii}, \& J.J. {Ostrowski}. 2016, \mnras,
  456, L45, \eprint{1506.05478}

\end{thebibliography}
\end{document}